# Data Science in Statistics Curricula: Preparing Students to "Think with Data"

J. Hardin, R. Hoerl, N. J. Horton, and D. Nolan

with: B. Baumer, O. Hall-Holt, P. Murrell, R. Peng, P. Roback, D. Temple Lang, and M.D. Ward

July 24, 2015

**ABSTRACT**

A growing number of students are completing undergraduate degrees in statistics and entering the workforce as data analysts.  In these positions, they are expected to understand how to utilize databases and other data warehouses, scrape data from Internet sources, program solutions to complex problems in multiple languages, and think algorithmically as well as statistically.  These data science topics have not traditionally been a major component of undergraduate programs in statistics.  Consequently, a curricular shift is needed to address additional learning outcomes.  The goal of this paper is to motivate the importance of data science proficiency and to provide examples and resources for instructors to implement data science in their own statistics curricula.   We provide case studies from seven institutions. These varied approaches to teaching data science demonstrate curricular innovations to address new needs.  Also included here are examples of assignments designed for courses that foster engagement of undergraduates with data and data science.

INTRODUCTION

The number of bachelor's degrees awarded in statistics has more than doubled in the five-year period 2008-2013 (Pierson, 2014) and continues to experience growth. This increase in the number of undergraduates may help address the impending shortage of quantitatively trained workers (National Academy of Sciences, 2010; Manyika, et al., 2011; Zorn, et al., 2014).  Statistics graduates at the bachelor's level often work as analysts, and as a result need training in statistical methods, statistical thinking and statistical practice; a foundation in theoretical statistics; increased skills in computing and data-related technologies; and the ability to communicate (ASA, 2014).   Computing skills to enable processing of large data sets are particularly relevant, as noted in the recent London Report on the Future of Statistics: "Undoubtedly the greatest challenge and opportunity that confronts today's statisticians is the rise of Big Data" (Madigan, 2014).

To help illustrate possible paths forward, we detail several existing approaches (many of which have been recently developed) to prepare students to work in industry as a statistician or an analyst, to continue their training in statistics graduate programs, or to become a scientist in a field allied to statistics.   Our goal is to provide resources and concrete ideas to the statistics community for how to incorporate computational and authentic data experiences into the undergraduate course work. To this end, we invited faculty from seven institutions (Johns Hopkins University, Purdue University, St. Olaf College, Smith College, University of Auckland, and the Universities of California, Berkeley and Davis) to describe their efforts in incorporating data science into the undergraduate curriculum in innovative ways.

#### WHAT IS DATA SCIENCE?

The term *data science* was suggested as a discipline by Cleveland (2001), who argued that the statistics profession should change its name to "data science", as that was, in fact, what statisticians did. Since then, the term data science has become a phrase describing a discipline typically involving some mixture of statistics and large-scale computing (Greenhouse, 2013).   Multiple definitions of a data scientist exist, including an inquisitive data explorer who communicates informed conclusions; someone who can use data from multiple sources to spot trends; or a "peculiar blend of developer and statistician that is capable of turning data into awesome" (Wills, 2015). We find particularly relevant the definition proposed by the NSF's advisory committee, StatSNSF, that data science is *the computational aspects of carrying out a complete data analysis, including acquisition, management, and analysis of data* (Johnstone & Roberts, 2014).

Given the vast increase in the volume and complexity of data and the new technologies that have been developed to process and analyze this information, we argue there is an increased need for statistical thinking in the context of working with data.  Key statistical reasoning topics that are critical for data scientists to know at a deep level include:
- Understanding the randomness, variability, and uncertainty inherent in the problem.
- Developing clear statements of the problem / scientific research question; understanding the purpose of the answer.
- Ensuring acquisition of high-quality data and not just a lot of numbers.
- Understanding the process that produced the data, to provide proper context for analysis.
- Allowing domain (subject-matter) knowledge of the problem to guide both data collection and analysis.
- Approaching modeling as a process that requires an overall strategy, not simply a collection of special techniques or methods.

**Why does data science belong in the undergraduate curriculum?**

We recognize many strengths of the current undergraduate curriculum in statistics.  It has at its core an understanding of variability.  Our students also learn that sampling biases can destroy the quality of inferential conclusions, and attendant issues of causation.  Issues regarding the responsible use of statistics and the proper conduct of science (e.g., issues of multiple comparisons and possible error rates) are imbued in many of our courses and programs.  In addition, simple statistical solutions to real problems are typically not complete without domain understanding and contextual knowledge.

Many statistical experimental designs, models, and analyses presented in a typical undergraduate statistics curriculum are necessary, but they are not sufficient for today's data problems.  To be relevant and able to tackle new data-rich problems, as well as the ever-increasing data challenges for the future, our students' statistical problem solving skills must be enhanced to incorporate practical computational skills.  The strong connections between statistical concepts and the scientific method provide our students with the ability to contribute in helping to frame and answer questions involving data.  What is not clear is whether they have equivalent ability to "think with data" (as enunciated by Diane Lambert of Google when a panelist on a Project INGenIOuS technology panel, http://www.maa.org/programs/faculty-and-departments/ingenious, D. Lambert, personal communication).  There is still a long way to go to ensure that they can tackle – both statistically and computationally -- issues with non-standard and non-textbook data.

According to Cuny et al. (2010), computational thinking involves "the thought processes involved in formulating problems and their solutions so that the solutions are represented in a form that can effectively be carried out by an information-processing agent."   As statisticians our first inclination is to focus on the inferential questions instead of the computational aspects.  However, as we encounter more and more complicated questions, data structures, and algorithms, the computational and algorithmic aspects become an integral part of arriving at a principled and statistically sound solution.  For example, the data problem could involve analysis of news media headlines via unstructured text scraped from the Web, microarray data extractions specified through HTML forms, or air traffic delays accessible from a Restful Web service.  The current statistical curriculum does not help address many of the obstacles we encounter in approaching these problems.  A number of questions might arise:  How do we find the data?  How do we retrieve the data? How can we assess the quality of the data?  What variables do we need to derive from the raw data?  How can the data be organized into a structure for analysis? Will the approach remain computationally feasible as the number of documents increases?

Nolan and Temple Lang's paper on "Computing in the Statistics Curricula" (2010) described new computational topics and skills that need to be incorporated within the statistics curriculum (at both the graduate and undergraduate levels).  They assert that computing skills are essential for scientific research, especially for scientific research containing statistical analyses, and that this will be increasingly true in the future.  While there has been some movement in this realm since the publication of their paper (see for example the American Statistical Association's guidelines for master's programs in statistics (Bailer, et al., 2012)), more work is needed.  To help encourage additional curricular ferment, we next provide several illustrations as to how the recommendations of Nolan and Temple Lang could be implemented at the undergraduate level.

**Integrating Data Science into the Curriculum: Seven Prototypes**

In order to illustrate how to develop novel data science curricula, we surveyed a number of faculty about their approaches to integrating data science into the statistics curricula. The descriptions presented here represent a number of innovative approaches. They are similar in that they all share a goal of having students become proficient in data technologies and programming tools for problem solving with data. However, their approaches vary in terms of mode of delivery, topics, and learning outcomes. Our goal in providing these examples is that these "existence proofs" can be useful to those who are working to integrate data science approaches into their own statistics curriculum.

The instructors have also shared their syllabi and some course materials in order to provide more concrete guidance about the types of modules, units, and assignments for others to adapt and adopt. The materials include lecture notes, class projects, homework assignments, etc. These resources are noted at the end of each prototype description. The syllabi have all been collected on the website, `http://hardin47.github.io/DataSciStatsMaterials/`. The data science exemplars are presented alphabetically by author.

**Data Science,** Ben Baumer, *Smith College*

The Data Science course at Smith College—first offered in 2013—is an elective in the new program in Statistical & Data Sciences and in the Applied Statistics minor. The course provides a practical foundation for students to compute with data, by participating in the entire data analysis cycle (from forming a statistical question, data acquisition, cleaning, transforming, modeling and interpretation). The course introduces students to tools for data management, storage and manipulation that are common in data science, and students apply those tools to real scenarios. Students undertake practical analyses using real, large, messy datasets using modern computing tools (e.g., R, SQL) and learn to think statistically in approaching all of the aspects of data analysis (see Baumer (2015) for a complete discussion of the course).

While some of the topics covered in the course come from existing offerings in applied statistics and computer science, effort is made in Data Science to present a unified blend of this material, such that students recognize that both fields contribute to answering questions from data. The course itself can be thought of as having five components: data visualization (e.g., data graphics, elements of visual perception), data manipulation (e.g., SQL, merging, aggregating and iterating), computational statistics (e.g., confidence intervals via the bootstrap, simulation, regression, variable selection), data mining/machine learning (e.g., classification, cross-validation), and additional topics (e.g., text mining, mapping, regular expressions, network science, MapReduce).

In its first offering, the staff of Smith's GIS (geographical information systems) laboratory regularly attended the class. This facilitated incorporation of lessons on spatial data and mapping techniques into the curriculum. This topic was popular with the students, since the ability to generate data maps was perceived to be useful in terms of visualization and communication.

A key learning outcome stressed in the course was the ability to think structurally about data and how to manipulate it.  Wickham's "key idioms" (2015) were used to illustrate for students the similarities and differences between merging and aggregating operations in R and SQL.  For example, what is the R equivalent of the GROUP BY operation in SQL? The SQL syntax is similar to the English language and so can help to demystify the R code. One key aspect of the course is helping students recognize that while each language may have its own syntax, the underlying operation that is being performed on the data is the same.

Among the biggest challenges is keeping students with varied computational abilities and backgrounds on the same page. For example, some students come in with extensive knowledge and practice with R, while others are seeing it for the first time. It is difficult to find assignments that keep both types of students motivated. Another challenge is maintaining a consistent level of difficulty and workload when cobbling together material from a variety of sources.

The end of course evaluations indicated that students felt as though they were learning things that are useful.  In turn, they have generated more interest among younger students in taking the course. Finally, several students have indicated that skills they learned in the course corresponded directly to questions that they were asked by employers during job interviews.

**Data Technologies,** Paul Murrell, *University of Auckland*

The Data Technologies course—first taught in 2002—introduces a variety of computer technologies relevant to storing, managing, and processing data (other courses in the Statistics Department at the University of Auckland teach more advanced programming and computational tools). Students with interests in applying statistics in business or research environments have found the course very useful, and enrollments have been steady at around 80 students for several years.

The course has two aims: to teach programming tools specific to the handling of data, and to teach and build confidence with general concepts of computer languages. Data Technologies also aims to build students' awareness of the range of tasks that a computer is capable of performing (in addition to providing concrete tools for performing specific tasks). Specific topics include: How to write computer code; publishing data on the World Wide Web (HTML); data description and semantic markup (XML); data storage (file formats, spreadsheets, databases); data management and summary (database queries, SQL); data processing (R).

The topics that are covered in this course are quite diverse, and there is very little overlap with other courses in the department.  This reflects the fact that everything leading up to the analysis of the data (data access, data cleaning) was not included in the curriculum.  The focus on R as a programming language also distinguishes this course from other courses that use R as a data analysis tool. This is a positive aspect to the course, because it reflects the fact that Data Technologies provides the students with a firmer basis of understanding and mastering of the language and computational approaches compared to those other courses.

Data Technologies has been perceived as being very rewarding for the students because they generally recognize how useful the course content is for their future, both inside and outside the university.  In other words, they know it is good for them, which has not necessarily been the case in other undergraduate courses in statistics.  Additionally, this course has lead to numerous students—who have continued on to do graduate work—seeking out the instructor for supervision of graduate projects.

The course has also been quite innovative in terms of delivery of course content on the Web and student submissions via the Web (other courses have generally shifted in that direction in recent years). However, the course remains quite innovative in its use of automated marking procedures to assist in assessment (taking advantage of the fact that student submissions are electronic and largely consist of computer code that can be evaluated).

One challenge the instructors of the course continue to face relates to formal assessment. The tests and the exams are paper-based, which is highly artificial, but very necessary to offset the ease with which electronic submission of computer-based coursework can be copied and shared between students.  The arrangement is awkward for both the students and the evaluators.  For the students, it is awkward because they are used to working on a computer, and for the instructor because, for example, code submissions cannot be run easily.  Having a computer-based examination environment would help, but would probably introduce headaches of its own.

Another challenge is that many students have only ever used a GUI interface like Microsoft Windows so there are some very strong mental models that have to be broken down and replaced.  For example, the notion that data files are distinct from executable files; that a file can be opened by applications other than the one that automatically opens when you double-click a file icon; that you are not limited by the menu options presented to you by a GUI interface, etc.  This situation continues to be a challenge with the growing prevalence of mobile devices where the code which runs the software is hidden behind barriers put up by the operating system.

All lecture, lab, and assignment material is available online (including lecture recordings). The course text is a CC-NC-SA licensed work, and is available at https://www.stat.auckland.ac.nz/~paul/ItDT/ (Murrell, 2009)**.**

**Concepts in Computing with Data,** Deborah Nolan, *University of California, Berkeley* and Duncan Temple Lang, *University of California, Davis*

These courses were co-developed at UC Berkeley and Davis in 2004.  It is a required upper division course for the statistics major in both departments and typically taken by sophomores and juniors. However, the majority of students enrolled in the course are not statistics majors. In 2014-15, over 600 students enrolled in four offerings of the course at Berkeley and over 200 students enrolled in one offering at Davis.

Both courses focus on the computational aspects of the data analysis cycle, from data acquisition and cleaning to data organization and analysis, and reporting.  Students are exposed to many different forms of data including structured data such as XML and JSON, free formatted text data, and dates, times, and geo-locations. To handle the data, students learn various tools and technologies including shell commands, regular expressions, structured query language (SQL) for relational databases, JavaScript for developing interactive Web pages, and R. Programming concepts are taught with R, including control flow, recursion, data structures and trees. Although the main focus is on the computational aspects of working with data, the course also covers many statistical topics, including concepts of variability, patterns, comparisons; exploratory and presentation graphics; statistical methods that are often not covered until late in the major program, such as classification trees, multi-dimensional scaling, and nearest neighbor methods; model selection and validation; and simulation tools, e.g., Monte Carlo, bootstrap, cross-validation.

One of the biggest challenges has been in developing the resources for projects and assignments.  This is more difficult than finding good data sets for teaching statistical methods because the data published for a statistical analysis often come already processed and cleaned, and the courses call for sources that are at least one step earlier in the analysis process.

Another common difficulty results from heterogeneity in student background. Many of the students are new to programming, but a substantial fraction of them have taken one or more CS courses. The first group is sometimes intimidated by the course.  For this reason, at Berkeley, graphics is taught first since students get very excited about the sophisticated plots that they can make with R.  At this point they are more open to learning programming concepts and handling more complex data (e.g., other than CSV formatted).  On the other hand, the more computationally advanced students often write code that works but ignores the paradigm of the language (e.g., loops instead of vectorized computations).  For these students, their programs are frustratingly slow. The challenge is to have them re-learn how to program with a different computational model.

Despite the challenges, Concepts in Computing with Data has been very rewarding to teach.  Several faculty have taught the course at Berkeley, and they all report how much they enjoy teaching it.  One of the rewards is seeing the enthusiasm that the students have for the material.   The confidence the students gain is noticeable as they tackle increasingly challenging computational problems.  They report how their project helped them get a job, gave them the confidence to learn new technologies in their new careers, and enabled them to participate in research projects in their major.  Many faculty routinely require this course for undergraduate students who wish to join their research team.

Additionally, Computing with Data enables teaching traditional statistical topics from a different approach.  For example, generating random numbers and carrying out simulation studies help students understand the concept of a random variable and its properties.  Also, students use exploratory data analysis (EDA) to debug code; cross-validation and bootstrapping to assess models and variability; and presentation graphics to summarize findings from advanced statistical analyses.

A dozen case studies from these and similar courses are now available (Nolan & Temple Lang, 2015a).

**Data Science Specialization,** Roger Peng, *Johns Hopkins Bloomberg School of Public Health / Coursera*

Roger Peng, along with his colleagues Jeff Leek and Brian Caffo, taught data science courses at Johns Hopkins in the MS Biostatistics program for many years. After realizing both the need for expanded access and the ease in which they could promote their courses, they took the set of data science courses to Coursera, a massive online open course (MOOC) platform. This material is now taught in nine modules as part of the Data Science Specialization. These modules are accessible and engaging to students with a large range of backgrounds (from those with very little background in statistics or computer science to those with PhDs in statistics or computer science who have not kept up with the data science movement). Each module runs for about 4 weeks so the entire program is 36 weeks long. Students can join the course at any of a variety of entry points depending on their background or desired educational goals.

The specialization provides instruction in the following areas: (1) The Data Scientist's Toolbox; (2) R Programming; (3) Getting and Cleaning Data; (4) Exploratory Data Analysis; (5) Reproducible Research; (6) Statistical Inference; (7) Regression Models; (8) Practical Machine Learning; and (9) Developing Data Products. The curricula of the modules cover the entire data science pipeline: formulating a context-relevant question or hypothesis, identifying sources of data and obtaining a data set, processing and transforming data so that it is suitable for analysis, summarizing and analyzing data to create statistical evidence, and finally creating and communicating statistical results in written format. Content covers the basics of building models based on new data types, experimental design, and statistical inference. Additional topics include statistical computing, reproducible research, data visualization, and machine learning.

The Data Science Specialization primarily uses the R programming language and focuses on its application in a variety of data analysis and statistical problems. Shell scripting and using various application programming interfaces (APIs) to access different kinds of data are also introduced (the use of APIs is discussed primarily in the context of accessing them from R via various packages). The expectations for students going into the course series is that they have some programming experience in another language.

The Data Science Specialization is taught entirely online; all assignments are graded by peers or machines. There are 3 types of assignments: programming assignments; quizzes (multiple choice, short answer), and peer assessments, which may include a variety of types of problems. Because the assignments are graded automatically, students receive instant feedback. Programming assignments are graded via unit tests so that R code is given specific inputs, and outputs are compared against correct output stored on the server. For peer assessments a minimum of four students must grade an individual's assignment in order for it to receive a grade. Conversely, students cannot view their own grades until they have graded a minimum of four assignments. Peer assessments have a grading rubric designed by the course instructors that details how points should be allocated to different aspects of the assignment. The benefits of peer assessment are that the course can scale up to large numbers of students and that students can learn from seeing other approaches to a problem.

One of the great benefits of the course series is the enormous reach via the MOOC format. Through various MOOCs to date, Jeff Leek, Brian Caffo and Roger Peng have taught nearly 1 million students. Being able to expose so many students at such a low cost is deeply rewarding. Teaching individual courses to such a large group of students (20k-30k per class) presents quite a few challenges. Primarily, the heterogeneity in the students' backgrounds is very high and the course material must be designed so that students with less background can "catch up" while students with more background are not bored. Designing a course in this manner is quite different from teaching the usual in-person class, even for large introductory lectures. Another key challenge is the inability to personally monitor the progress of each student and to assist them when needed.

Leek, Caffo, and Peng have developed software tools (see swirl, http://swirlstats.com) to address the pacing challenge by allowing students to learn certain technical material at their own speed. The R package -- swirl -- aims to teach users statistics and R simultaneously and interactively. It attempts to create an authentic learning environment by guiding users through interactive lessons directly within the R console. The software presents a choice of course lessons and interactively tutors a user through them. A user may be asked to watch a video, to answer a multiple-choice or fill-in-the-blanks question, or to enter a command in the R console precisely as if he or she were using R in practice. Emphasis is on the last aspect, interacting with the R console. User responses are tested for correctness and hints are given if appropriate. Progress is automatically saved so that a user may quit at any time and later resume without losing work. Once a lesson is completed, the Coursera Web site is notified so that the student can receive credit for completing the lesson.

The entire course series is presented online through the Coursera platform (https://www.coursera.org/specialization/jhudatascience/1). Lectures are presented in video format with embedded video quizzes.

**A Statistics-infused Introduction to Computer Science,** Paul Roback and Olaf Hall-Holt, *St. Olaf College*

At St. Olaf College a data-driven (or statistics-infused) introduction to computer science for budding statisticians and others who need to understand and use computational tools in handling data was developed as part of a collaboration between the Statistics and Computer Science Programs. The course, called CS125 Computer Science for Scientists and Mathematicians, was first offered in spring 2013. It is a recommended elective for the statistics concentration at St. Olaf, while also serving as a prerequisite for courses in the computer science major as well as a way for students to satisfy their general education requirement in abstract and quantitative reasoning. In its three initial offerings, CS125 has been bursting at the seams, drawing from diverse audiences that span second-semester seniors seeking a deeper understanding of R after seeing it in several previous statistics courses, to students attracted to future statistics courses after their exposure to data-centric projects, and students recognizing the course as potentially useful for future scientific research. The course description appears below:

> CS125 introduces popular tools for handling data, including 1) obtaining data from Web sources, 2) visualizing the data, 3) searching for important patterns in multidimensional data, and 4) sharing results on the Web. The emphasis in this course is not just a particular set of tools, but also a broad perspective on computing that applies to many different application areas over time. The approach is hands-on, working with real world datasets. Students in biology, chemistry, mathematics, psychology and statistics are especially encouraged to consider this course.  The only prerequisite is Calculus I; no prior experience with statistics or programming is required.

Students completing CS125 should understand fundamental programming principles—computational and algorithmic thinking as a problem solving strategy—and have established a base of knowledge from which they can easily pick up new tools in the future.  With this in mind, the development team decided to introduce two languages: R and Python.  Python and R are both widely used by data scientists, often in tandem, since they have different strengths and weaknesses.  Python is a great beginning language with a rich ecosystem of pedagogical materials and is often convenient for data collection and formatting.  R is used in all courses in the statistics concentration at St. Olaf for data analysis and visualization. Students can also benefit from a language comparison, whereby they can glimpse the difference between a foundational computational structure and a choice that is specific to a single programming language.  The statistics faculty was highly supportive of the decision to provide the students with the tools to use both Python and R.  Generally, the faculty hope to enable statistics students to be more self-sufficient in collecting and preparing data, especially data that does not immediately follow traditional rows-as-observational units and columns-as-variables formatting (e.g., unstructured text).

The course is taught in an environment conducive to interactive learning. The classroom is equipped with 16 Linux machines and 9 projectors connected to these machines.  Before each class, students are expected to read selections from assigned texts (and sometimes search the Web for certain answers and approaches) and apply the ideas they've read about to several exercises.  The first part of each class is devoted to a discussion of questions that students submitted before class, while the second half of class often features small groups of students at the boards working on in-class problems designed to illustrate coding logic.

In some ways, CS125 at St. Olaf is a work in progress, but in other ways it has already successfully filled a strong need.  The biggest challenges are the heavy workload for an introductory course and the diversity of students attracted to the course.  Students willing to invest the necessary time report high satisfaction with the knowledge they accumulate; however, the current set of topics and assignments will probably have to be pared down to avoid turning off students who are newer to computing or who are balancing heavy course loads, since some students bridle at the "teach yourself" ethos that is sometimes seen in many computer science courses.

For the statistics curriculum, this course has quickly become indispensable.  By far the most common suggestion statistics concentrators make during exit surveys and interviews is to require CS125 early in the concentration, or at least to advise students to take it at some point.  Since the courses in statistical modeling and theory are methods-focused, they can only devote so much time to the finer points of R programming and troubleshooting.  Junior and senior concentrators who have taken CS125 have universally praised the insights gained into R (and computing in general).

For faculty in statistics and computer science at St. Olaf, the development and implementation of CS125 has provided rewards that have grown from their collaborative approach.  The statistics faculty saw how computer scientists develop and connect ideas in programming, what tools are effective for data organization and scraping, and how to present these ideas through out-of-class and in-class work.  The computer science faculty learned R, which is used less often in their community, and saw how statisticians think about data and then use software to gain insights into their data.  By jointly offering the course, both the statistics and CS faculty are better able to stay abreast of the changes that the data science movement is making to both disciplines.  More immediately, the computer scientists have a new course which supports a growing body of data-centric non-majors while providing potential majors a valuable start, and the statisticians have a popular elective course in beginning data science for their concentrators without reducing course offerings elsewhere.  As they speak with colleagues at other institutions, the St. Olaf faculty are struck by the perceived barriers to collaboration between many statisticians and their computer scientists colleagues.  This model demonstrates the value of engaging each other in conversations and initiatives that can be enriching for both faculty members and students.

All daily assignments and larger projects can be found on the class wiki at http://www.cs.stolaf.edu/wiki/index.php/CS125.

**An Introduction to Big Data Analysis**, Mark Daniel Ward, *Purdue University*

Purdue University is one of the largest producers of undergraduate statistics majors.  The Introduction to Big Data Analysis course is part of a new initiative in Purdue's NSF-funded Statistics Living-Learning Community (STAT-LLC), which began in fall 2014.  The STAT-LLC is a unique, immersive experience for approximately 20 students per year.  It unites many of the elements of the undergraduate experience. The program is aimed at sophomores, with the goal of creating a bridge from the first-year general curriculum into sophomore year Statistics major courses and into a student's first research experience in data analysis, especially with big data. Through this unique experience, the expectation is for students to be:
- more likely to stay in their chosen major program (improved retention rates),
- more confident in their coursework and their research,
- more successful during their sophomore year academic courses, and
- well-positioned for graduate school and post-graduate experiences and careers.

The program includes academic courses, residential life, professional development, and mentored research projects that last a full calendar year (as opposed to a summer research experience).  The students take three core courses as a cohort: probability theory, statistical theory, and this new course in big data analysis.  They live together in a common residence hall with a dining court.  They also enroll in a year-long professional development seminar that touches on all aspects of university life, and on their future careers and training. Their research experiences are supported by faculty mentors from statistics and applied disciplines.

The Introduction to Big Data Analysis course focuses on computational tools for representing, extracting, manipulating, interpreting, transforming, and visualizing data. It is a statistics elective and has no prerequisites. The first half of the course introduces students to the R platform, and to basic data structures, exploratory data analysis, data visualization, random number generation and simulation, and an introduction to linear models. The second half of the course includes topics such as bash shell and shell scripting; awk, regular expressions, and pattern matching; relational databases using SQL; and XML parsing and Web scraping. In the future, it is likely that the course will be expanded to include parallelism and distributed data with Hadoop and MapReduce.

The course is taught in a flipped environment. The course webpage contains videos, computer code, and notes about the topics. The entire course is project-based, using data sets chosen from various areas of application. Students work in teams in a computer laboratory and perform all of their computations remotely, on a server. Through the projects and assignments, the students gain practical experience in effectively communicating insights about data.

All course materials are available at http://llc.stat.purdue.edu/2014/29000/index.html.

**CURRICULAR TOPICS**

The prototypes described in the previous section differ in level and audience. They range from introductory courses in data technology, to a core course for the undergraduate major, to a 36-week data science specialization. Yet, many topics arose repeatedly, and for faculty who have just started to consider how to introduce data science in their curricula, Table 1 may provide ideas for key topics. Decisions about what topics to include, how much time to dedicate to them, and in what sequence they should be covered will in part depend on curricular constraints and decisions of the local institution. Despite this, we provide here some ideas to consider when developing and updating curriculum to include data science topics.

*Programming.* Programming is an essential skill for data science. As seen in Table 1, we consider programming to include concepts of structured programming and higher order notions of efficiency, and in some cases, high performance computing. It is no longer adequate training for statistics students to be able to analyze data using graphical user interfaces or to write simple scripts that do not use modular approaches (including writing functions and code using control flow) to process data. At some institutions, it may be possible to have programming experience as a prerequisite for a data science course. However, most of the example courses here do not have such a prerequisite. This is predominately due to three constraints: the extra prerequisite limits enrollment because students in some fields, e.g., the social sciences, might not have taken the prerequisite course; requiring such a course may be perceived by these same students as a barrier to enrollment; and the data science course offers an important and unique integration of programming concepts with data handling and analysis. This integration yields a course that has a very different focus from typical introductory programming courses and that can ameliorate the difficulties with learning programming concepts because they are couched in a framework that emphasizes learning from data.

*Data technologies and formats:* These two areas are essential to any data science course. The division of these topics into the two areas of technologies and format in Table 1 is somewhat arbitrary. For example, a relational database has a specific data format but it is included under technologies because SQL is a language for accessing data in a database. Similarly, text data appears under the format heading, but it typically requires some level of familiarity with regular expressions to extract information for analysis. The University of Auckland is the exception in the courses presented in not addressing text and regular expressions. This is due to the focus of that course on Web technologies and because the material is covered in one of the other two computing electives in their program. The topics of XML and shell commands are not universally included in these courses. There are many arguments for including and excluding these topics in the course. For example, knowledge of the shell is very useful for programmatic handling of files, such as thousands of twitter messages. It also has the advantage of offering an example of trees and hierarchical data structures, an important CS concept that can help students understand how information is organized.

*Statistical Topics:* We believe it is crucial to include statistical topics in the course and not simply limit topics to those of data wrangling because an understanding of how we might analyze the data impacts how we process the data. Additionally, as many of the course creators have indicated, they see data science as offering an opportunity to expose students to the entire data analysis process and to teach statistical thinking in alternative, more realistic contexts. Some of the prototypes introduce visualization early in the course. Two reasons are given for this: it provides a platform for introducing computational notions that lead to rewarding outcomes, e.g., a beautiful plot, and because visualization can be reinforced throughout the rest of the course. Including modern methods in the curriculum provides fun ways to apply more complex and advanced methods. In such a course, the focus tends to be more on the statistical ideas and less on formal properties of the methods. Many advanced methods would be appropriate to include, with the selection focusing on those that are computationally intensive and intuitive. Examples include recursive partitioning, support vector machines, nearest-neighbor methods, little bag of bootstraps, and LASSO. The topic of simulation can be approached from a resampling perspective and/or a simulation study/experiment. The collection of these topics seems more an artifact of the presence or absence of other computational courses in the curriculum. For example, at Auckland there are three computing courses and so less pressure to include "everything." Additionally, since the writing of this article, UC Davis has converted their one-quarter computing course into three quarter courses. The first focuses on programming and includes workflow and simulation topics. The second has a focus on data technologies, and the last covers new topics on advanced computing methods, e.g., high performance computing.

| Area | Topic | Smith | Auckland | UC B/D | JHSPH | St. Olaf | Purdue |

|              |                 | # Weeks | 14 | 12 | 14/10 | 36 | 14 | 15 |
|--------------|-----------------|---------|----|----|-------|----|----|----|
| Programming  | Structured      |         | X  | X  | X     | X  | X  | X  |
|              | Efficiency      |         |    |    | X     | X  | X  | X  |
|              | HPC             |         | X  |    |       |    |    |    |
| Data Tech    | RDBMS (SQL)     |         | X  | X  | X     | X  | X  | X  |
|              | RegEx           |         | X  |    | X     | X  | X  | X  |
|              | XML             |         |    | X  | X     | X  |    | X  |
|              | Shell commands  |         |    |    | X     | X  |    | X  |
|              | Web scraping    |         | X  |    | X     | X  | X  | X  |
| Data Formats | Ragged arrays   |         |    | X  | X     | X  | X  | X  |
|              | Text data       |         | X  |    | X     | X  | X  | X  |
|              | Data cleaning   |         | X  | X  | X     | X  | X  | X  |
| Work Flow    | Reproducibility |         | X  |    |       | X  |    |    |
|              | Web publishing  |         |    | X  | X     | X  | X  |    |
|              | Revision control|         |    |    |       | X  |    |    |
| Statistical  | Simulations     |         | X  |    | X     | X  | X  | X  |
|              | Modern method   |         | X  |    | X     | X  | X  | X  |
|              | Visualization   |         | X  |    | X     | X  | X  | X  |

*Table 1. Curricular topics offered in each of the example courses. See the definitions section for more detailed explanations of the topics.*

*Definitions of computation-oriented terms.*

*Structured Programming* **(Structured)**: a programming paradigm that uses conditional statements, iteration (e.g., for and while loops), block structures, and subroutines.
*Efficiency*: the speed of runtime execution of code.
*High Performance Computing* **(HPC)**: techniques of parallel processing and grid computing that use many computing resources simultaneously.
*Relational Databases* **(RDBMS (SQL))**: information stored in multiple tables (i.e., relations). Each table represents an entity, with rows (records) and columns (variables), and allows linking between tables.
*Regular Expressions* **(RegEx)**: a language for describing patterns to search for in text.
*eXtensible Markup Language (XML)*: a text-based format for exchanging information. XML obeys a set of rules for encoding documents that is human readable and machine readable and generated.
*Shell commands*: A command-line interface to the operating system's file and process management.
*Web scraping:* Automated procedures for retrieving content from the Web.
*Ragged arrays*: non-rectangular data where records have differing numbers of values.
*Reproducibility*: the notion that a final product includes the computations required to produce the results, such as code, data, computing environment, etc.
*Revision Control*: software to manage collaborative development, editing, and sharing of code, documents, web sites, etc.

It is difficult to quantify the time spent in a course on the topics found in Table 1 because topics are often taught simultaneously and early topics are reinforced when covering later topics. Also, we did not include the basics of learning a computer language, e.g., expressions and data types, under the programming topic in Table 1.  However, we attempt to give the reader a better sense of the balance of topics for two courses (St Olaf and Berkeley) by examining the assignments for the courses. At St. Olaf, the student work consists of 29 homework assignments due at each class meeting and 5 larger projects.  At Berkeley the student work consists of 11 weekly lab assignments, 8 homework assignments, and 2 projects (plus a midterm and final).  For each course, we reviewed the various assignments and attempted to categorize them according to the topics in Table 1. When assignments covered multiple topics we distributed the work evenly across the relevant topics.  We hope this crude estimate gives a sense of the distribution of time/effort across the computational topics for these two courses.  For each type of assignment, we list the topics in order from greatest to least and provide percentages for those that are more than 10% of the total.  We note that the nature and philosophy of the assignments is not captured by this metric.  See below for examples

At St. Olaf, the daily assignments fell into the following categories:  nearly 75% of the work focused on programming concepts in R and Python (including the basics mentioned above that are not listed in the table). The remaining topics were, in decreasing order, regular expressions, Web scraping, visualization, web publishing, and SQL.  The balance of topics for the five projects was quite different. Programming made up about 50% of the project focus, followed by modern methods (20%), visualization (15%), simulation, regular expressions and web publishing.

At Berkeley about 35% of the lab work focused on programming, and 10% each was on regular expressions, SQL, XML shell, simulation, and visualization.  For the homework, again about 35% was on programming. After that, the general focus was on statistical topics with about 25% on visualization, 15% on modern methods, and, in decreasing order, simulation, regular expressions, SQL, and XML. For the projects, the effort on the topics is divided as follows: programming 30%, visualization 30%, simulation 15%, modern methods 15%, web publishing, regular expressions.  We also quantified the lecture time as follows: programming 35%, visualization 15%, modern methods 15%, then simulation, regular expressions, Web scraping, SQL, shell, XML, shell, and Web publishing.

**Examples of Assignments and Projects**

Next, we turn our attention to assignments and projects which can help to bring data science topics into the curriculum, through four specific examples of assignments from the data science courses.  The resources for these assignments are also available on the git repository.

Baumer developed an assignment for his data science course that investigated the claim that every actor in Hollywood is connected to Kevin Bacon by no more than 6 degrees of separation; a similar but more substantive analysis from quantitative sociology might involve the use of network analysis on observed interactions between drug-involved individuals in a community (Johnson, et al., 2013).  Individuals who are most central to the network are likely good targets for information and inquiry into the problem. For the course assignment, first the students figure out how to address the research question; then they use SQL to gather and organize the relevant data from the IMDb database (URL). Next, they address the research question, and last they learn about measures of centrality (e.g., betweenness centrality and eigenvector centrality, similar to Google page rank). Other assignments help students see the strengths and weaknesses as well as advantages and disadvantages of different tools.

Nolan and Temple Lang have students develop a spam filter for email. The data consist of over 9,000 email messages that have been classified by SpamAssassin (http://spamassassin.apache.org). The first step is to process the raw text messages, which appear in individual files, into a form conducive to statistical analysis. For example, students use regular expressions to separate the header from the body and attachments and extract from the header the key-value pairs, e.g., the sender, time it was sent, subject line, etc., and create variables and values containing this information. The students also use regular expressions to locate the words in the email body and convert them into a word vector, i.e., counts of occurrences of all words in the corpus. After converting the email into information that can be used for analysis, the students, e.g., use naïve Bayes to calculate the likelihood a message is spam given its word vector. Alternatively, students write functions to derive variables from characteristics of a message and use these variables to classify email with a decision tree. In both approaches, the assignment continues with the students choosing a threshold/model via cross-validation, assessing the fitted model with test data that they have set aside in advance, and considering the confusion matrix of false positives and negatives.

One team assignment from the St. Olaf course is called The "Residence Hall Energy Use" project. This project asks students to visualize on-campus energy use data and then apply those techniques to energy usage on a global scale. This assignment, however, differs from a typical statistics course project by emphasizing the entire process of data analysis and the computational aspects of each step. Initially, students clean the data by writing functions to manage categorical variables, calculate continuous summaries, and split strings. Students additionally use random forests to impute missing data, and ggplot2 (Wickham, 2009) to visualize the data. Finally, students must interpret their plots at both the campus and the global levels.

Peng and collaborators have coincidentally embraced the communication aspects related to statistics and data science that the ASA's new undergraduate curriculum guidelines (http://www.amstat.org/education/curriculumguidelines.cfm) emphasize. In particular, one of the courses within the specialization is on Developing Data Products, which encompasses the formal presentation of a statistical analysis via an interactive Web interface. The students learn to use the Shiny application in R (http://shiny.rstudio.com) to create interactive graphics and communicate uncertainty in statistical results. The final project for this module in the Data Science Specialization has students create a Shiny Web application (see examples, http://shiny.rstudio.com/gallery/) with associated documentation. Additionally, each student pitches their product using reproducible R computations which is both embedded in the Shiny application, presented using Slidify, and pushed to github.

SUMMARY DISCUSSION

As faculty who have worked to integrate aspects of data science into our own curricula, we have a number of reflections on the various implementations presented here.  The most striking highlight is that all faculty report how popular their courses are and how rewarding they are to teach.  Not surprisingly, a course that incorporates data science can be a way to excite students about further study in statistics.  We conjecture that one reason for high enrollments in the data science classes at these institutions is that the students perceive these courses as relevant and exciting.

We also contend that incorporating data science into the statistics curricula gives us an opportunity to teach statistical methods, thinking, practice, and computation in a modern venue that is meaningful to students and aligned with the goals of a statistics education.  van der Laan (2015)  defined statistics as the science of learning from data.   We assert that to analyze the data of today and the future these techniques need include to include more data related skills.

In the data science courses described above, students have the opportunity to repeatedly participate in the entire data analysis cycle: forming a research question, obtaining data, formatting & cleaning data, analyzing, and communicating results.   Whereas the first few aspects of carrying out an analysis have typically been skipped in too many traditional courses and curricula (Zhu, et al., 2013) , using a data centric approach requires students to follow the entire trajectory in order to understand the source and inherent variability of the data. Moreover, with a data science course it is possible to integrate statistical thinking with computing with experience with data -- the goal is to focus on the computational problem solving aspects of carrying out a data analysis.  The recommendation for repeated exposure to this cycle flows directly  from guidance on the structure of the statistics curriculum (ASA, 2014).

Class projects and assignments in data science courses differ from traditional computer science assignments and from traditional data analysis assignments because they typically require both a demonstration of computing aptitude and insight into the data analysis.  Grading of the assignments can be cumbersome. Systems such as Murrell's relieve the burden of grading code. Assignments such as these often require the students to write up their analysis and findings, and Peng's peer grading system tries to address this aspect of the grading.  Both of these grading systems are useful in reducing the burden of grading when there are a large number of students in the course.  However, there are additional considerations that need to be addressed.  Statistics faculty have typically not been trained to evaluate technical writing or coding of their students, which in turn means the students (i.e., peer graders) do not receive training either.  Moreover, assignments that require data analysis are often very open-ended and developing tests for code can be difficult without making the assignment overly prescriptive.   For faculty who are interested in developing their pedagogical abilities in data science, we recommend participation in relevant workshops and seminars that focus on such topics.  Recent workshops include the 3$^{rd}$ Computation and Visualization Consortium Workshop (http://www.macalester.edu/hhmi/curricularinnovation/cvc/) and NSF Workshops through the UC Davis Data Sciences Initiatives (http://datascience.ucdavis.edu/NSFWorkshops/).

It is worth noting that the courses described here have very few, if any, prerequisites. Some require introductory statistics, and one has a computer science requirement. The lack of requirements may be one reason for the heterogeneity in student background, which all of the faculty we surveyed noted as a challenge. We also think that it is due to limited exposure to computing concepts in K-12 education. Despite this disparity, the faculty have devised a variety of approaches to keep the students engaged and challenged but not overwhelmed. For example, they have developed assignments that can be adapted or extended to meet the disparate background levels of the students. Other strategies employed include the careful sequencing of material, creating learning communities through peer learning cohorts or peer grading, and modular assignments.

The course implementations also provide examples of how data science can be successfully integrated into the statistics curricula. Faculty considering implementing a data science course must face the question of how the new course fits into their curriculum. Berkeley and Davis require the data science course for the major, but Auckland, Smith, and St. Olaf offer it as an elective. We note that Berkeley reduced the number of mathematics requirements from 5 to 4 in order to add this new requirement. Davis initially offered the course as an elective, and soon afterwards, the value of this training for advanced courses in statistical methods was recognized among the faculty and it became a required course. Alternatively, St. Olaf has successfully developed a mutually beneficial partnership between the Statistics and CS programs, which enables them to offer this course through CS and not statistics. Collaborations, in general, may offer a model for other schools where faculty workload is a constraint or where only a limited number of statistics courses are offered. Although the course in data technologies is not required at Auckland, the department offers several popular electives in computing.

We further claim that the topics covered in data science courses can and should be brought into a variety of statistics courses. That is, although we have presented full classes on data science, our hope is that many of the examples, assignments, and topics presented in Table 1 make their way into other statistics courses. Key data science skills need to be introduced, reiterated, and reinforced throughout the statistics curriculum. For example, it is possible to bring small examples of SQL queries into introductory statistics (Horton, et al., 2015). Early introduction of data science ideas and methods will not only make the tasks in the later data science course easier, but it may excite students about the wealth of exciting topics within statistics – encouraging them to go on to take a data science course (Gould, 2010). After having taken a data science elective, capstones, summer experiences (e.g., Summer Institute for Training in Biostatistics, Explorations in Statistical Research (Nolan & Temple Lang, 2015b)) and competitions can solidify skills garnered in a data science course. For example, a senior thesis or capstone project that incorporates data management and manipulation, scraping data from the Web, and interactive data visualizations, allows students to solidify their data skills in a way that a typical course cannot.

Competitions allow for students to put their abilities to use. Recent contests have played a central role in getting students excited about statistics and exposing them to authentic data analysis problems. For example, see the DataFest (Gould, et al., 2014), Kaggle predictive modeling competitions

([https://www.kaggle.com/competitions](https://www.kaggle.com/competitions)), ASA's Data Expo (http://stat-computing.org/dataexpo/), Undergraduate Statistics Project Competition (USPROC, https://www.causeweb.org/usproc/), and Midwest Undergraduate Data Analytics Competition (MUDAC, http://www.mudac.org/). It is important to note that these competitions take the onus off the faculty member *teaching* an entire course and changes it into a student run/lead investigation solving a real data problem. Additionally, students are able to display their data science skills publicly. Indeed, at the 2014 and 2015 Five College DataFest competition, a team of students from Baumer's class won Best in Show for their analyses.

In summary, the main objective of this paper is in making concrete recommendations about ways that new capacities in data science can be implemented in the undergraduate curriculum. Nolan & Temple Lang(2010), Brown & Kass (2009), ASA (2014), Cobb (2015) and others have called for a comprehensive restructuring of how students are prepared to deal with the myriad of data they will see in their careers. This paper takes these recommendations a step further by offering a variety of example implementations, along with syllabi and course materials to compare and contrast, to adopt and adapt, and to assist faculty who want to modernize their statistics programs.

ACKNOWLEDGEMENTS

Partial support for this work was made available by NSF grant 0920350 (Project MOSAIC, [http://www.mosaic-web.org](http://www.mosaic-web.org)). We would also like to acknowledge and thank the referees who provided valuable feedback and suggestions for improvement on the manuscript.
ACKNOWLEDGEMENTS

Partial support for this work was made available by NSF grant 0920350 (Project MOSAIC, [http://www.mosaic-web.org](http://www.mosaic-web.org)). We would also like to acknowledge and thank the referees who provided valuable feedback and suggestions for improvement on the manuscript.